\documentclass[prd,letterpaper,11pt,twoside,tightenlines,nofootinbib,showpacs,preprint,twocolumn]{revtex4}
\usepackage{graphicx}
\usepackage{latexsym}
\usepackage{epsfig}
\usepackage[english]{babel}
\usepackage{graphicx}
\usepackage[T1]{fontenc}
\usepackage[utf8]{inputenc}
\usepackage{amsmath}
\usepackage{amssymb}

\usepackage{bbold}

\begin{document}
\newcommand{\eg}{{\it e.g.}}
\newcommand{\etal}{{\it et. al.}}
\newcommand{\ie}{{\it i.e.}}
\newcommand{\be}{\begin{equation}}
\newcommand{\dd}{\displaystyle}
\newcommand{\ee}{\end{equation}}
\newcommand{\bea}{\begin{eqnarray}}
\newcommand{\eea}{\end{eqnarray}}
\newcommand{\bef}{\begin{figure}}
\newcommand{\eef}{\end{figure}}
\newcommand{\bce}{\begin{center}}
\newcommand{\ece}{\end{center}}

\title{Thermodynamic Geometry of Nambu- Jona Lasinio model}
\author{P. Castorina$^{2,4}$, D. Lanteri$^{1,2}$, S. Mancani$^3$}
\affiliation{
\mbox{${}^1$ Dipartimento di Fisica, Universit\`a di Catania, Via Santa Sofia 64,
I-95123 Catania, Italy.}\\
\mbox{${}^2$ INFN, Sezione di Catania, I-95123 Catania, Italy.}\\
\mbox{${}^3$ Dipartimento di Fisica,  Universit\`a di Roma “La Sapienza”, Piazzale Aldo Moro 2, 00185 Roma, Italy}\\
\mbox{${}^4$ Institute of Particle and Nuclear Physics, Faculty of Mathematics and Physics, Charles University}\\
\mbox{V Hole\v{s}ovi\v{c}k\'ach 2, 18000 Prague 8, Czech Republic}\\
}

\begin{abstract}

The formalism of Riemannian geometry is applied  to study the phase transitions in Nambu -Jona Lasinio (NJL) model.
Thermodynamic geometry reliably describes the phase diagram, both in the chiral limit and for finite quark masses.  
The comparison between the geometrical study of NJL model and of (2+1) Quantum Chromodynamics at high temperature and small baryon density  shows a  clear connection between chiral symmetry restoration/breaking and deconfinement/confinement regimes.

\end{abstract}
 \maketitle

\section{Introduction}

Geometry, and in particular differential geometry, is now considered a powerful tool to study statistical systems.

Indeed, information geometry \cite{info1,info2,info3}, which started with the seminal paper by Rao~\cite{Rao} has emerged from studies of invariant geometrical
structure involved in statistical inference. It defines a Riemannian metric together
with dually coupled affine connections in a manifold of probability distributions.

These geometric structures play important roles not only in statistical inference but also in
wider areas of information sciences, such as machine learning, signal processing,
optimization, neuroscience, mathematics and, of course, physics \cite{info1,info2,info3}.

Thermodynamic geometry (TG),  a specific application of information geometry methods to equilibrium thermodynamics, started with an initial \cite{Rao,Wein1975} definition  of 
a metric for statistical systems,  i.e. a measure of the ``distance'' between different thermal equilibrium  configurations, later  refined in ref.~\cite{Ruppeiner:1979} by determining the metric tensor, $g_{\mu \nu}$, through the Hessian of the entropy density. 

This definition of  $g_{\mu \nu}$ is crucial since the resulting distance is in inverse relation with the fluctuation probability between equilibrium states and, moreover,  it leads to
the ``interaction hypothesis'', i.e the correspondence between the absolute value of the scalar curvature $R$ (an intensive variable, with units of a volume, evaluated by the metric) and  $\xi^3$, the cube  of  the correlation length, $\xi$,  of the thermodynamic system. Indeed,  a covariant and consistent thermodynamic fluctuation theory can be developed~\cite{Ruppeiner:1995zz}, which  generalizes the classical fluctuations theory and offers a theoretical justification to the physical meaning of $R$.

TG  has been tested in many different systems: in phase coexistence for Helium, Hydrogen, Neon and Argon ~\cite{Ruppeiner:2011gm}, for the Lennard-Jones fluids~\cite{May:2012,May:2013}, for ferromagnetic systems and liquid-liquid phase transitions~\cite{Dey:2011cs};  in the liquid-gas like first order phase transition in dyonic charged AdS black hole~\cite{Chaturvedi:2014vpa}; in the Hawking-Page transitions in Gauss-Bonnet-AdS black holes~\cite{Sahay:2017hlq}. 

More recently~\cite{,Castorina:2018ayy,Castorina:2018gsx}, TD  has been applied to field theories and, in particular, to  Quantum-Chromodynamics (QCD) at large temperature and low baryon density, to evaluate the \mbox{(pseudo-)~critical} deconfinement temperature $T_c$  and to  compare the results with the Hadron Resonance Gas models.

In this paper  a systematic  application of TD to the Nambu - Jona Lasinio (NJL) model is carried out. This study is not only interesting per se, since the NJL model gives clear indications on some dynamical mechanism, as chiral symmetry, for low energy QCD  but also because a QCD fundamental property, quark confinement, is missing in NJL model with some interesting  consequences on the geometrical description.

The TD approach is recalled in Sec.~\ref{sec:TG} and in Sec.~\ref{sec:NJL}  the phase diagram of the Nambu-Jona Lasinio model is discussed.
Sec.~\ref{sec:NJLG} is devoted to the thermodynamic geometry description of chiral symmetry restoration in NJL model in the chiral limit and for finite fermion masses.
The geometrical difference in describing  QCD and NJL phase transitions is considered in Sec.~\ref{sec:QCD} and Sec.~\ref{sec:CC} contains our comments and conclusions.

\section{Thermodynamic Geometry \label{sec:TG}}

In this section the procedure to define the thermodynamic metric is briefly recalled (the details are in ref. \cite{Ruppeiner:1995zz,Ruppeiner:1998}) and the description of phase transitions by the scalar curvature, $R$, is discussed, making also use of the application to real fluids. 
\subsection{Thermodynamic metric}

Let $A_U$ be a large thermodynamic system (universe) and let us  consider an open subsystem $A$ with thermodynamic coordinates $a^0$, the internal energy density, and $a^i$, the number densities of particles of different species. 
The probability density to find $A$ in the ``point'' $a=(a^0,a^1,\cdots)$ is given by
\begin{equation}\label{eq:Probability}
P(a,a_U)\,d^na =C\;e^{S_U(a,a_U)}\;d^na\;,
\end{equation}
being $C$ a normalization constant, \mbox{$a_U=(a^0_U,a^i_U,\cdots)$} denotes the state of the universe and $S_U$ its total entropy, formally regarded as an exact function of the parameters of $A$ and $A_U$. 
   
On the basis of the maximum entropy principle and
in the framework of Consistent and Covariant Fluctuation Theory (CCFT)~\cite{Ruppeiner:1995zz}, 
the thermodynamic properties of $A$ can be studied through the introduction of a quadratic form,
\begin{equation}\label{eq:dist}
\left(\Delta \ell\right)^2 = g_{\mu\nu} \;\Delta a^\mu \; \Delta a^\nu 
\;,
\end{equation}
where $\Delta a^\mu = a^\mu - a^\mu_U$ and
\begin{equation}\label{eq:g2}
g_{\mu\nu} = - \frac{\partial^2 s}{\partial a^\mu \partial a^\nu}\Bigg|_{a=a_U }
\end{equation}
defines a positive-definite Riemannian metric on the space of thermodynamic states as the Hessian of the entropy density, $s$, with respect its natural variables $a^\mu$.

One can show~\cite{Ruppeiner:1995zz} that previous formulas give the probability of the spontaneous fluctuations between equilibrium states.
Indeed, by expanding eq.~\eqref{eq:Probability} up to second order for $a \simeq a_U$, the maximum entropy state, one finds the classical gaussian normalized fluctuation probability density:
\begin{equation}\label{eq:PG}
\begin{split}
P(a,a_U)\,d^na = &\left(\frac{V}{2\,\pi} \right)^{\frac{n}{2}}
\sqrt{g_U}\;\times\\
&\times\;\exp\left\{-\frac{V}{2}\,g_{\mu\nu}\,\Delta a^\mu \,\Delta a^\nu\right\}\,d^na\;,
\end{split}
\end{equation}
being $g$ the determinant of $g_{\mu\nu}$  and $\sqrt{g}\,d^n x$ the usual invariant volume on a Riemannian manifold.

In the analysis of the phase transitions in NJL model by thermodynamic geometry  we shall consider  a two dimensional manifold, where the intensive coordinates are $\beta=1/T$ and $\gamma = -\mu/T$, with $\mu$ chemical potential. Moreover the metric~\eqref{eq:g2} turns out to be  related with the derivatives of the potential $\phi=p/T$, where $p$ is the pressure~\cite{Ruppeiner:1998}:
\begin{equation}
g_{\mu\nu} = \left(\begin{array}{cc}
\phi_{,\beta\beta} & \phi_{,\beta\gamma}\\
\phi_{,\beta\gamma} & \phi_{,\gamma\gamma}
\end{array}\right)
\;,
\end{equation}
with the usual comma notation for derivatives. 

The scalar curvature $R$ simply becomes
\begin{equation}
R
 =
 \frac{1}{2\,g^2}\; 
  \left|\begin{array}{ccc}
\phi_{,\beta\beta} & \phi_{,\beta\gamma} & \phi_{,\gamma\gamma}\\
\phi_{,\beta\beta\beta} & \phi_{,\beta\beta\gamma} & \phi_{,\beta\gamma\gamma}\\
\phi_{,\beta\beta\gamma} & \phi_{,\beta\gamma\gamma} & \phi_{,\gamma\gamma\gamma}
\end{array}\right|
\;.
\end{equation}

\subsection{Phase transition in thermodynamic geometry} 

The main results of the thermodynamic geometry within Ruppeiner's formulation~\cite{Ruppeiner:1995zz} are: 1)~the (inverse) relation between
the line element and the fluctuation probability between equilibrium states; 2)~the, so called, \textit{Interaction hypothesis}: the absolute value of the scalar curvature $R$ is proportional to a power of the correlation length, i.e.  $|R|\sim\xi^d$, where $d$ is the effective spatial dimension of the underling thermodynamic system. 
  
\begin{figure}
	\centering
	\includegraphics[width=0.6\columnwidth]{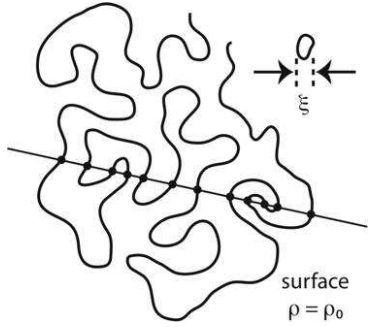}
	\caption{Schematic picture of the meaning of $\xi$: the intricate line represents the surface of $\rho(r)=\rho_0$, i.e. that separating two sides with local mean densities $\rho>\rho_0$ and $\rho<\rho_0$. By tracing any straight line, the intersection points are separated by an average distance equal to $\xi$. Figure~from~\cite{Ruppeiner:2012}.}
	\label{fig:xi} 
\end{figure}

\begin{figure}
	\centering
	\includegraphics[width=0.75\columnwidth]{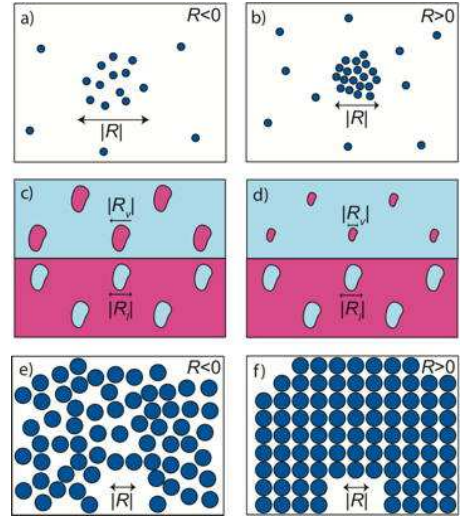}
	\caption{Schematic pictures of different possible particle arrangements: (a) cluster of particles with volume $|R|$ pulled together by the attractive part of the interparticle interaction ($R<0$); (b) a repulsive solid-like cluster held up by hard-core particle repulsion ($R>0$); (c-d) a fluid in two phases near the critical point: the bottom half is a liquid phase containing vapor droplets with volume $|R_l|$. The top half is a coexisting vapor phase containing liquid droplets with volume $|R_v|$. In (c) $|R_v|=|R_l|$ and the droplets are commensurate, in (d) liquid and vapor phases have incommensurate 	droplets; 
	(e) liquid phase; (f) solid phase with $R>0$.
	Figure~from~\cite{Ruppeiner:2012}.}
	\label{fig:Rvari} 
\end{figure}

The meaning of the correlation length and of the scalar curvature can be represented as in Fig.\ref{fig:xi} (a schematic picture due to Widom~\cite{Widom:1974}): the intricate line represents what the surface of density $\rho(r)=\rho_0$ might look at any instant. This surface separates two sides with local mean densities $\rho>\rho_0$ and $\rho<\rho_0$. By tracing any straight line, the intersection points with the surface $\rho_0$ are separated by an average distance equal to $\xi$. 	 Because such  points are separated by the same mean distance $\xi$,  whatever the direction of the line,  it is convenient to think that regions as volume elements (``droplets'') of dimension $R\sim \xi^d$.
Figure~\ref{fig:Rvari} shows a schematic summary of different  configurations.

The interaction hypothesis has been confirmed by the study of the classical ideal gas ($R=0$~\cite{Ruppeiner:1979}) and of the van der Waals gas~\cite{Ruppeiner:1995zz}, for which, near the liquid-vapor critical point, $T_c$, the curvature is $R\sim \left|\left(T-T_c\right)/T_c\right|^{-2}$. 

Other confirmations come from the study of the Takahashi Gas~\cite{Ruppeiner:1995zz}, the Curie-Weiss model~\cite{Janyszek:1989}, the ferromagnetic monodimensional Ising
model ~\cite{Janyszek:1990}. For a more complete list of applications see  Tab.~I of Ref.~\cite{Ruppeiner:2010}.

The relation between $|R|$ and $\xi^d$ is easy to verify for second-order phase transitions, since $R$ diverges, but the criterium to define a new phase in term of the curvature $R$ for a first order phase transition or a crossover is less clear.

The approach called \textit{$R$-Crossing Method} (RCM) ~\cite{Ruppeiner:2011gm} is often applied to define first order phase transitions. It is  based on the continuity  of the scalar curvature:  knowing the thermodynamic quantities in the two phases, i.e. $R$, one can build up the transition curve by imposing the continuity of $R$. 
The RCM, coherent with  Widom’s microscopic description of the liquid-gas coexistence region (i.e. with the idea that the correlation lengths of the two phases must be the same at the transition)
has been tested in systems with different features: vapor-liquid coexistence line for the Lennard-Jones fluids~\cite{May:2012,May:2013}, first and second order phase transitions of mean-field Curie-Weiss model (ferromagnetic systems), liquid-liquid phase transitions~\cite{Dey:2011cs}, phase transitions of cosmological interest as the  liquid-gas-like first order phase transition in dyonic charged AdS black
hole~\cite{Chaturvedi:2014vpa}.
Another criterion, applied in the study of first order phase transitions in real fluids~\cite{Ruppeiner:2012} and Lennard-Jones systems~\cite{May:2013}  is  a first kind discontinuity in $R$. 

Finally, two different phases can be linked by a crossover, as for the QCD deconfinement transition. Also in this case there is no definitive conclusion on the behavior of $R$, although, 
 it has been recently  shown~\cite{Castorina:2018ayy} that the condition $R = 0$  predicts a  temperature for the transition from QCD to the Hadron resonance Gas at low baryon density
in agreement with freeze out curve~\cite{Floris:2014pta,Das:2014qca,Adamczyk:2017iwn} and (within $10\%$) with lattice data \cite{Steinbrecher:2018phh,Bazavov:2017dus}. 

Another interesting aspect of the geometrical approach to phase transitions is that the sign of the scalar curvature  brings information on the microscopic interactions, since $R$ turns out to be positive for fermi statistical interactions and negative in the bosonic case  ~\cite{Janyszek:1990b,Ubriaco:2016}.  Therefore a change in sign of  $R$  is an indication of the balance between effective interactions, even when no transition occurs, and theoretical curves with $R=0$ in pure fluids  identify  some anomalous behaviors observed in the experimental data of several substances (in particular, water)~\cite{Ruppeiner:2017,Ruppeiner:2012}. A transition from $R > 0$ to $R < 0$  has been also shown for the Lennard-Jones system~\cite{May:2013,May:2012} and Anyon gas~\cite{Mirza:2008fy,Ubriaco:2013}. For black holes~\cite{Sahay:2010tx}, the  the change in sign of the curvature occurs at the Hawking-Page transition temperature, therefore  associated with the condition $R = 0$.

In the next sections we shall apply the thermodynamic geometry approach to NJL phase diagram both in the chiral limit and for finite fermion mass.
The behavior of the scalar curvature in the quantitative description of the critical line in the $T - \mu$ plane will be pointed out.

\subsection{An example: real fluids} 

The geometrical study of fluids is based on the Helmholtz free energy per volume, $f$, in terms of  $(T,\rho)$ coordinates ($T$ is the temperature, $\rho = N/V$ is the particle density) and the corresponding thermodynamic line element is given by~\cite{Ruppeiner:2012}
\begin{equation}
\Delta \ell^2 = -\frac{1}{T}\left(\frac{\partial^2 f}{\partial T^2}\right)_\rho \Delta T^2
+
\frac{1}{T}\left(\frac{\partial^2 f}{\partial \rho^2}\right)_T \Delta \rho^2
\end{equation}
The scalar curvature turns out to be
\begin{equation}
R = \frac{1}{\sqrt{g}}\;\left[
\frac{\partial }{\partial T}\left(\frac{1}{\sqrt{g}}\;\frac{\partial g_{\rho\rho}}{\partial T}\right)
+
\frac{\partial }{\partial \rho}\left(\frac{1}{\sqrt{g}}\;\frac{\partial g_{TT}}{\partial \rho}\right)
\right] 
\;,
\end{equation}
with 
\begin{equation}
g_{TT} = -\frac{1}{T}\left(\frac{\partial^2 f }{\partial T^2} \right)_{\rho}
\;,\qquad
g_{\rho\rho} = \frac{1}{T}\left(\frac{\partial^2 f }{\partial \rho^2} \right)_{T} 
\;.
\end{equation}
and $g=g_{TT}\,g_{\rho\rho}$.

In Ref.~\cite{Ruppeiner:2012,Ruppeiner:2015,Ruppeiner:2017} the real fluid free energy is modeled on the NIST Chemistry WebBook and $R$ is evaluated in the liquid and vapor phases and along the liquid-vapor coexistence curve ending at the critical point $T_c$.

At the critical point $R  \rightarrow - \infty$  with a power law behavior and in the asymptotic critical region, i.e very close to the critical temperature, the values of the scalar curvature evaluated in the two phases coincide. However in other regions of the thermodynamic parameter space the values of $R$ in the liquid and the vapor phases~\cite{Ruppeiner:2012} are quite different and mesoscopic fluctuating structures of different sizes  occur in the two phases (see fig.~\ref{fig:Rvari}.d). 

In the  phase diagram of fluids,  $R$ is generally found to be negative since the average molecular distances are such that the attractive part of the intermolecular potential dominates. 
However different anomalous regions, i.e. with $R>0$,  exist (see fig.4 in ref. \cite{Ruppeiner:2017}). They are localized: (a) in the supercritical liquid region, near the melting line; (b) in the liquid phase near the triple point (for water); (c)  in the vapor phase, in some regions  called ``repulsive clusters''~\cite{Ruppeiner:2017}.
 
The thermodynamic states for cases (a) and (b), named solid-like-liquid states, emerge when the liquid organizes into solid-like structures at large densities, with a small intermolecular average separation.
The states in ``repulsive cluster'' areas (case c), are characterized by values of $R$ much larger than the volume of a single molecule and  by low density and have been observed in 97 different fluids (except those consisting of the simplest molecules) along the saturated vapor phase curve.

\section{ Nambu - Jona Lasinio Model}\label{sec:NJL}

In Nambu–Jona Lasinio (NJL) model with two flavors ($f=u,\;d$), the $SU(2)$  lagrangian~\cite{Klevansky:1994,Klevansky:1999,Buballa:2003qv} is given by 
\begin{equation}
\begin{split}
\mathcal L_{SU(2)}=&\overline \psi_f  \left(i\,\partial\!\!\!/-m\right)\psi_f +\\
& + G\,\left[\left(\overline \psi_f \psi_f\right)^2+\left(\overline \psi_f i \gamma_5 \overrightarrow \tau\psi_f\right)^2\right]\;,
\end{split}
\end{equation}
being $G$ a dimensionful coupling, $m$ the current quark mass ($m=0$ is the chiral limit) and $\overrightarrow \tau$ the Pauli matrices.
In mean-field approximation the thermodynamic potential, $\Omega$,  at finite temperature and chemical potential turns out to be~\cite{Buballa:2003qv}
	\begin{equation}\label{eq:Omega}
	\Omega(M_f) = 
	\frac{\left(M_f-m\right)^2}{4\,G}
	+N_f\;\Omega_f
	\;,
	\end{equation}
	with
	\begin{equation}\label{eq:OmegaF}
	\begin{split}	
	\Omega_f	=&
	-
	2\,N_c\int \frac{d^3 p}{(2\,\pi)^3}\;E_f  
	-\\
	&
	-
	2\,N_c\,T\int \frac{d^3 p}{(2\,\pi)^3}\;\ln\left[1+e^{-\textstyle{\frac{E_f+\mu_f}{T}}}\right] -\\
	&-
	2\,N_c\,T\int \frac{d^3 p}{(2\,\pi)^3}\;
	\left[1+e^{-\textstyle{\frac{E_f-\mu_f}{T}}}\right]
	\;,
	\end{split}
	\end{equation}
where $M_f$ is the dynamically generated mass, $E_f=\sqrt{p^2+M_f^2}$,  $N_c$ and $N_f$ are the number of colors and flavors respectively, $\mu_f$ is the quark $f$ chemical potential and the integrals are regulated by a cutoff $\Lambda$. For  $m_u=m_d$,   $\mu=\mu_u=\mu_d$, the generated quark mass is  $M=M_u=M_d$ .

To evaluate the minimum of $\Omega$ by eq.~\eqref{eq:Omega}, one has to solve the self-consistent gap equation
\begin{equation}\label{eq:GAP}
M = 
m-2\,G\,\left<\overline \psi \psi\right>\;,
\end{equation}
where $\left<\overline \psi \psi\right>$ is  the quark-antiquark condensate:
\begin{equation}
\begin{split}
\left<\overline \psi \psi\right>=
-2\,N_c\,N_f\!\!\int\!\!\!
\frac{d^3p}{(2\,\pi)^3}\,\frac{M}{E} \; \Psi(T,\mu)
\;,
\end{split}
\end{equation}
with
\begin{equation}\label{eq:Psi}
\Psi(T,\mu)=
1- n_+(\mu)
- n_-(\mu)
\end{equation}
and
\begin{equation}\label{eq:n}
n_\pm(\mu) = \frac{1}{1+\exp\left\{\frac{E\pm\mu}{T} \right\}} \;.
\end{equation}
For three flavors  with $m_u=m_d=m$, and $m_s\neq m$, one has $M_u=M_d\neq M_s$, and the $SU(3)$ lagrangian is~\cite{Buballa:2003qv}
\begin{equation}
\mathcal L_{SU(3)}=\overline \psi  \left(i\,\partial\!\!\!/-\widehat m\right)\psi +  \mathcal L_4+\mathcal L_6\;,
\end{equation}
where
\begin{equation}\label{eq:l4}
\mathcal L_4 = G\,\sum_a\left[\left(\overline \psi \lambda_a\,\psi\right)^2+\left(\overline \psi i \gamma_5 \lambda_a\psi\right)^2\right]
\end{equation}
and the  't Hooft interaction, $\mathcal L_6$, is given by
\begin{equation}
\mathcal L_6 = -K\Bigg[\det \overline \psi \left(1+\gamma_5\right)\psi 
+\det \overline \psi \left(1-\gamma_5\right)\psi\Bigg]\;,
\end{equation}
with $\psi = \left(u,d,s\right)^T$,
$\widehat m=\text{diag}\left(m,m,m_s\right)$, $\lambda_0 = \sqrt{2/3}\;\mathbb{1}_{3\times 3}$, being $\mathbb{1}_{3\times 3}$ the $3\times3$ identity matrix, 
 and where $\lambda_a$ ($a=1,\;\ldots,\;8$) are the Gell-Mann matrices and $K$ and $G$  dimensionful couplings.

The gap equations,  
\begin{equation}\label{eq:GAP3q}
\begin{split}
M_i = m_i -4\,G\,\left<\overline \psi_i\psi_i\right>
+2\,K\,&\left<\overline \psi_j\psi_j\right>\left<\overline \psi_k\psi_k\right>\\
&\qquad
(j,\;k\neq i)
\;,
\end{split}
\end{equation}
are coupled with the quark condensates
\begin{equation}\label{eq:c}
\begin{split}
\left<\overline \psi_i \psi_i\right>=
-2\,N_c\int
\frac{d^3p}{(2\,\pi)^3}\,\frac{M_i}{E_i} \; \Psi_i
\;,
\end{split}
\end{equation}
where
\begin{equation}
\Psi_i =1 - \frac{1}{1+e^{\frac{E_i+\mu_i}{T}}}
-
\frac{1}{1+e^{\frac{E_i-\mu_i}{T} }}
\;,
\end{equation}
and the mean-field thermodynamic potential $\Omega$ turns out to be ~\cite{Buballa:2003qv}
\begin{equation}\label{eq:Omega3q}
\begin{split}
\Omega = 
\sum_{f=u,d,s}\Omega_f &+ 2\,G\,\sum_{f=u,d,s}\left<\overline \psi_f \psi_f\right>^2-\\
& 
-4\,K\,\left<\overline u u\right>\left<\overline d d\right>\left<\overline s s\right>\;,
\end{split}
\end{equation}
with $\Omega_f$ in eq.~\eqref{eq:OmegaF}.

Finally, the potential we need for the thermodynamic geometry calculations is
\begin{equation}\label{eq:phi}
\phi(\beta,\gamma) = \frac{P}{T} = -\Omega(\beta,\gamma)\;\beta \;,
\end{equation}
where $P=-\Omega$ is the pressure.

\section{Thermodynamic geometry of chiral symmetry restoration in NJL model}\label{sec:NJLG}

\subsubsection{Two flavors in the chiral limit} \label{sec:NJL2chi}

Let us first discuss  the chiral limit ($m=0$) for two flavors, starting from the breaking of chiral symmetry at $T=\mu=0$, with 
the  value of the dynamical mass $M_0(0,0)=300\,MeV$, corresponding to $\Lambda=650$~MeV and $G=5.01\times 10^{-6}\;MeV^{-2}$~\cite{Klevansky:1994,Klevansky:1999}.

\begin{figure}
	\centering 
	\includegraphics[width=\columnwidth]{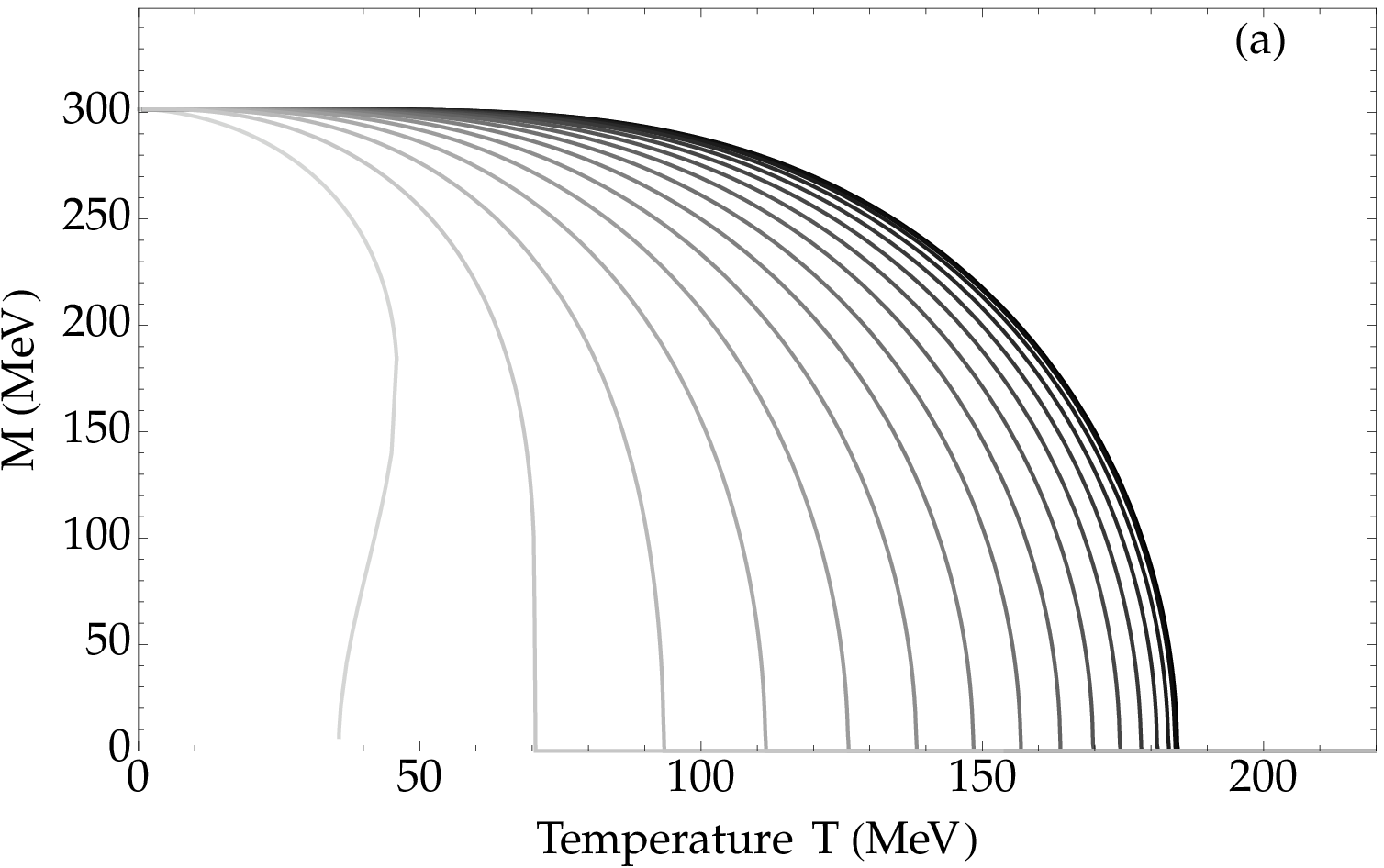}
	
	\includegraphics[width=1.02\columnwidth]{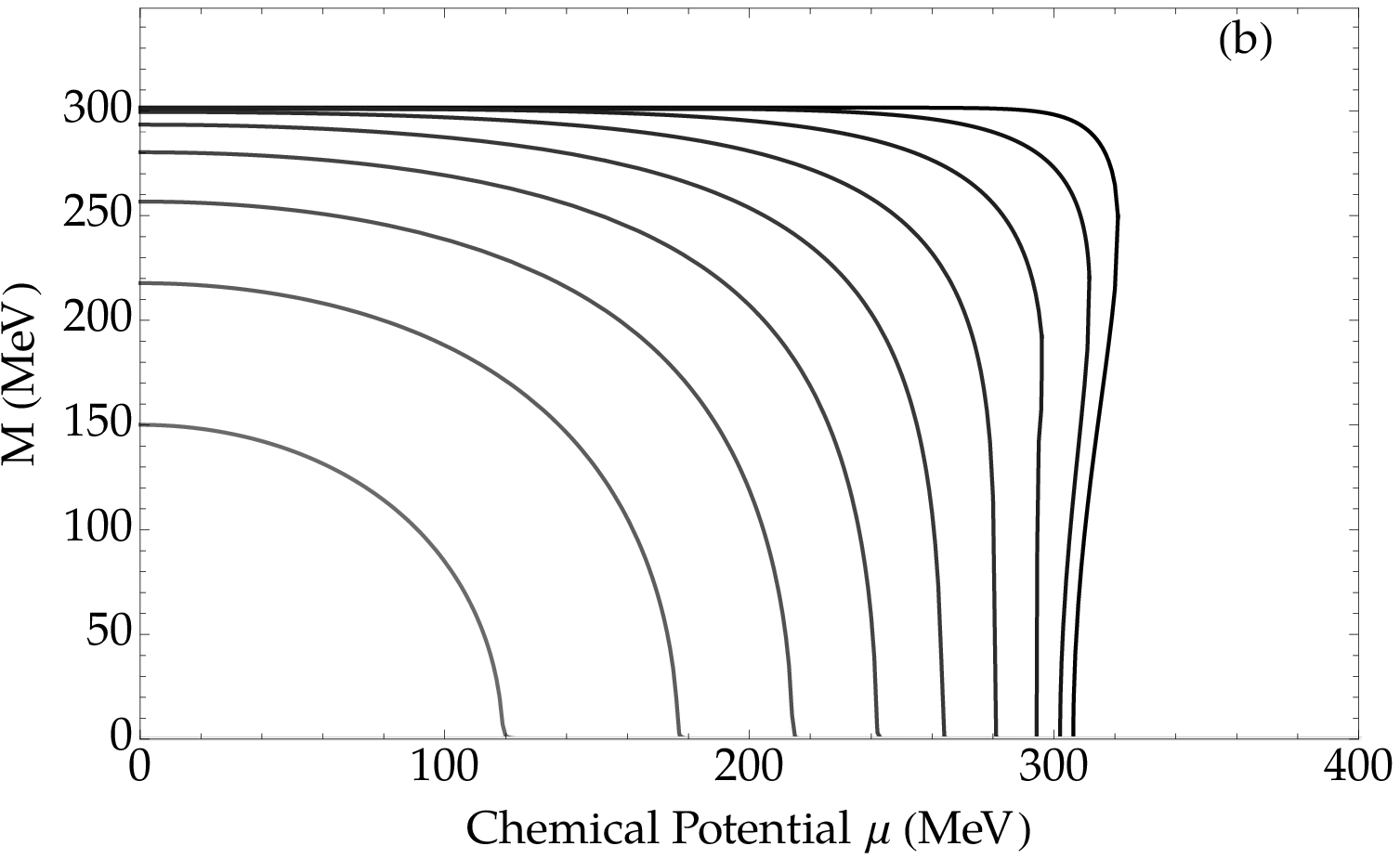}
	\caption{a) The dynamically generated mass, $M$, in the NJL model
		with two favors in the chiral limit ($m_u = m_d = 0$ MeV) and  again the temperature. Black line is for $\mu = 0$ MeV; the others are for growing $\mu$, up to $\mu = 300$ MeV and with step of $\Delta \mu = 20$ MeV. b)  $M$ as a function of the chemical potential $\mu$. Black line is for $T =10$ MeV; the others are for growing $T$, up to $T = 170$ MeV and with step of $\Delta T = 20$ MeV.}
	\label{fig:M2fChi}
\end{figure}

The well known solution $M(T,\mu)$  of the gap equation~\eqref{eq:GAP},  for different values of the  temperature and of the quark chemical potential,
 is plotted in Fig.~\ref{fig:M2fChi}.a and  \ref{fig:M2fChi}.b.  The restoration of the chiral symmetry is a first order phase transition  at large chemical potential and a second order one at low $\mu$.

The study of the critical line of the symmetry restoration,  $T(\mu)$,  by thermodynamic geometry requires the, straightforward but laborious,  calculation of the scalar curvature $R$, reported in appendix~\ref{app:NJL2f}.

\begin{figure}
	\centering 
	\includegraphics[width=1\columnwidth]{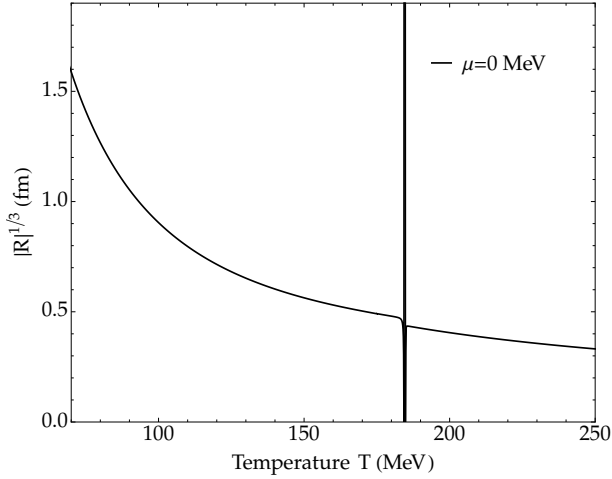}
	\caption{$R$ from $\mu=0$~MeV:  second order phase transition.}
\label{fig:R2fChiII}
\end{figure}

\begin{figure}
	\centering 
	\includegraphics[width=\columnwidth]{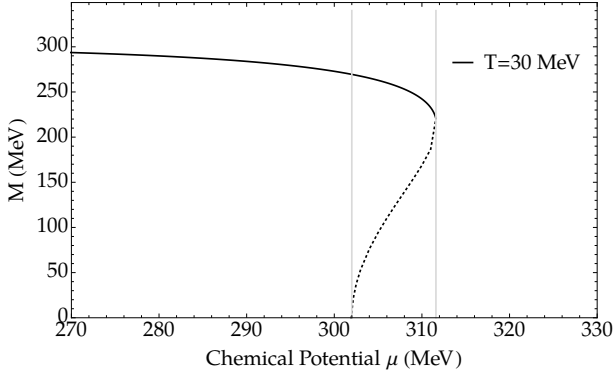}
	
	\caption{The dynamically generated mass $M$ in the 2 flavors NJL chiral model and temperature $T=30$~MeV.}
	\label{fig:MChiT30}
\end{figure}

\begin{figure}
	\centering 
	\includegraphics[width=\columnwidth]{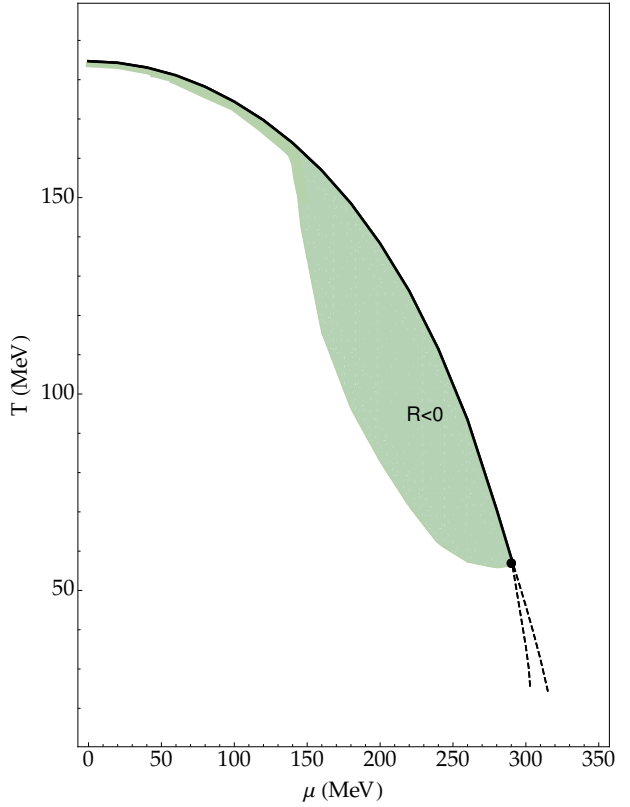}
	\caption{The transition temperature: continuous line is for II order phase transition and the dashed ones for the first order one. The transition point is at $\mu_\chi^\star = 290$~MeV and $T_\chi^\star=58$~MeV. The green band is the region of $R<0$}
	\label{fig:T2fChi}
\end{figure}

It turns out that  $|R|$ diverges at the critical temperature, i.e. there is a second order phase transition, for $\mu <\mu^\star \simeq 290$ MeV, as shown in fig.~\ref{fig:R2fChiII} for $\mu=0$. For
$\mu>\mu^\star$ there is, instead, a first order phase transition. The dynamically generated mass, $M$, now takes the characteristic behavior plotted in figure~\ref{fig:MChiT30}, where the black curves (both the continuous and the dotted) are for $T=30$~MeV and the two light-gray lines define the spinodal points. Between the two spinodal (light-gray)  lines one can evaluate three different scalar curvatures:  the first one  for the higher-mass branch (black curve in figure~\ref{fig:MChiT30}); the second one for $M=0$~MeV and the last one is related to the $M$-branch that interpolates between $M=0$ and the upper $M$-curve (dotted curve in figure~\ref{fig:MChiT30}). 
At fixed temperature and between the spinodal lines (see fig.~\ref{fig:MChiT30}), there is a discontinuity in $|R|$ which identifies the two dashed curves in fig.~\ref{fig:T2fChi}.

The crossing temperature from the I order phase transition to the II order turns out to be about $ 58$~MeV.

For small $\mu$ and near the transition the curvature is negative, i.e. the interaction is mostly attractive, suggesting that the chiral symmetry restoration is  due to thermal fluctuations.

On the other hand,  at large chemical potential $R$ turns out to be positive, indicating a screening of the potential and an increase of the repulsive interaction at large density.

The complete critical line obtained by thermodynamic geometry is depicted in  Figure~\ref{fig:T2fChi} where the continuous line shows the II order phase transition and the dashed lines the spinodal curves of the first order one. The green band is the region of negative $R$.

\subsubsection{Two flavors with chiral masses}

With finite chiral quark masses, at high temperature and low chemical potential, there is  a smooth crossover rather than a second-order phase transition. Moreover,  the first-order phase
boundary ends in a second-order endpoint~\cite{Buballa:2003qv}.

\begin{figure}
	\centering 
\includegraphics[width=\columnwidth]{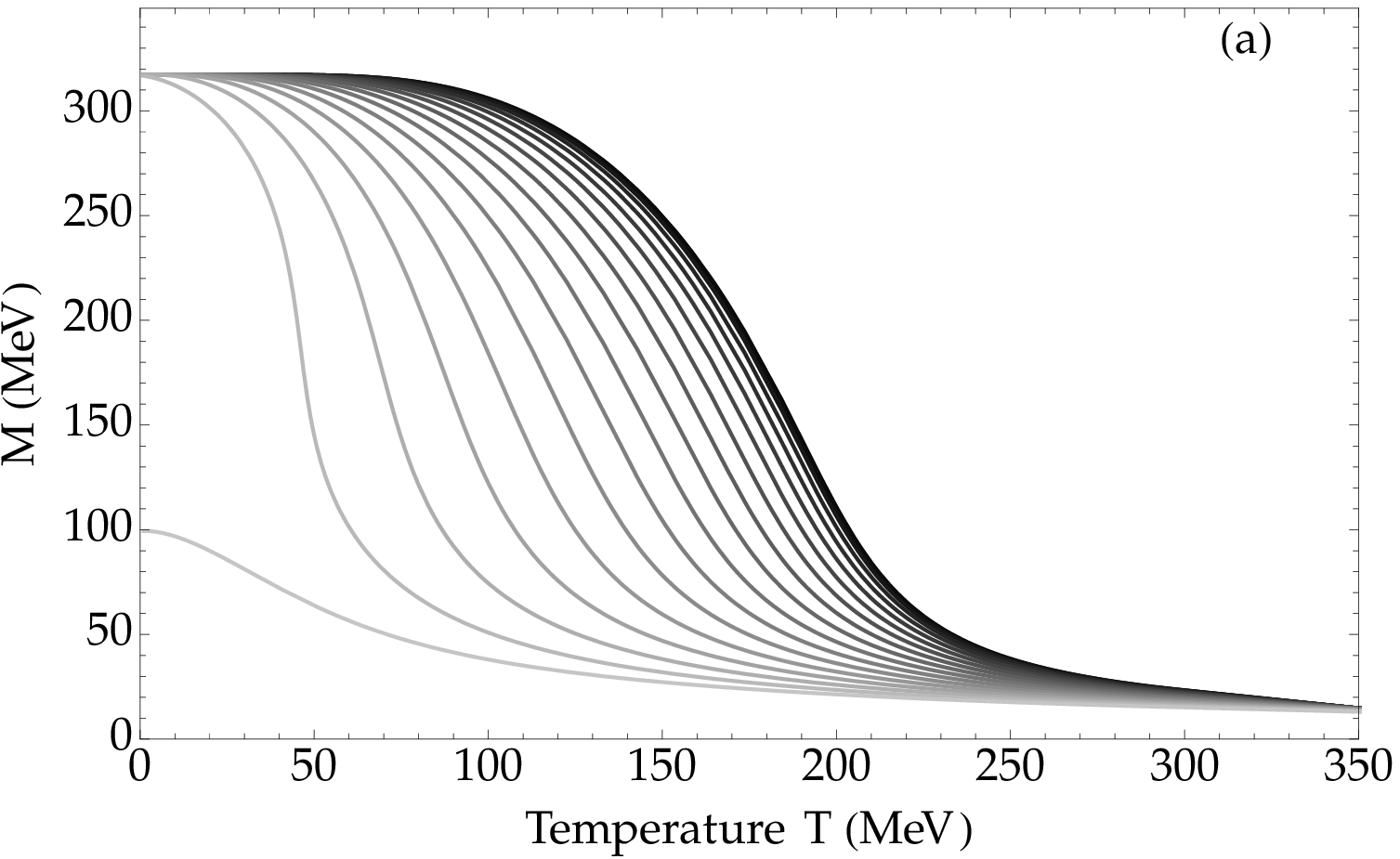}

\includegraphics[width=0.96\columnwidth]{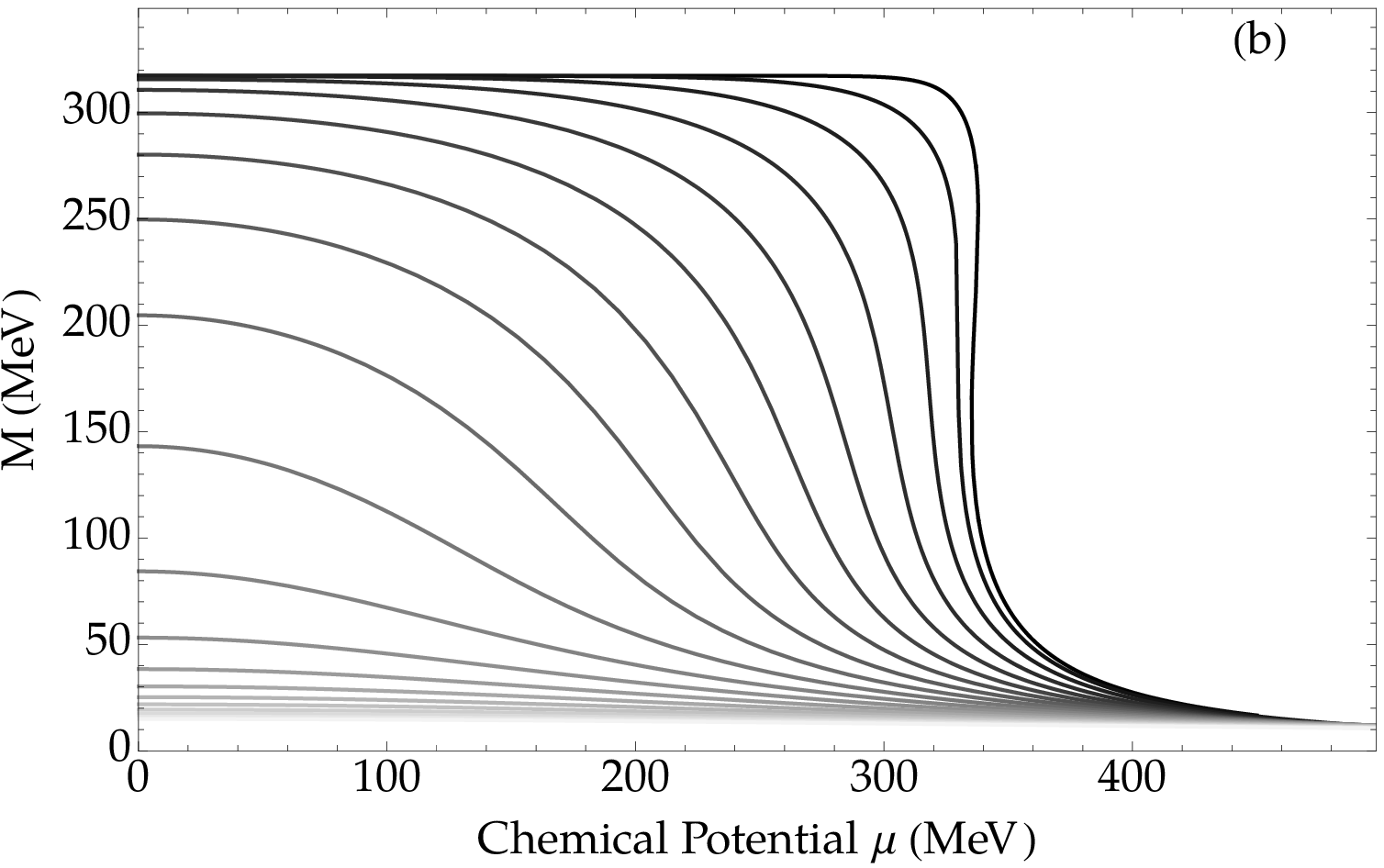}
		\caption{a) The dynamical generated mass, $M$, in the NJL model
			with two favors of identical mass ($m_u = m_d= 5.5$ MeV) and  again the temperature. Black line is for $\mu = 0$ MeV; the others are for growing $\mu$, up to $\mu = 340$ MeV and with step of $\Delta \mu = 20$ MeV. b)  $M$ as a function of the chemical potential $\mu$. Black line is for $T =10$ MeV; the others are for growing $T$, up to $T = 400$ MeV and with step of $\Delta T = 20$ MeV. 	 }
		\label{fig:M2f}
\end{figure}

The solution of the  gap equation~\eqref{eq:GAP} (with $\Lambda=650$~MeV and $G=5.01\times 10^{-6}\;\text{MeV}^{-2}$ and $m_0=5.5$~MeV)
as a function of $T$ and $\mu$ is shown in Figs.~\ref{fig:M2f}.a and \ref{fig:M2f}.b.

\begin{figure}
	\centering 
	\includegraphics[width=1\columnwidth]{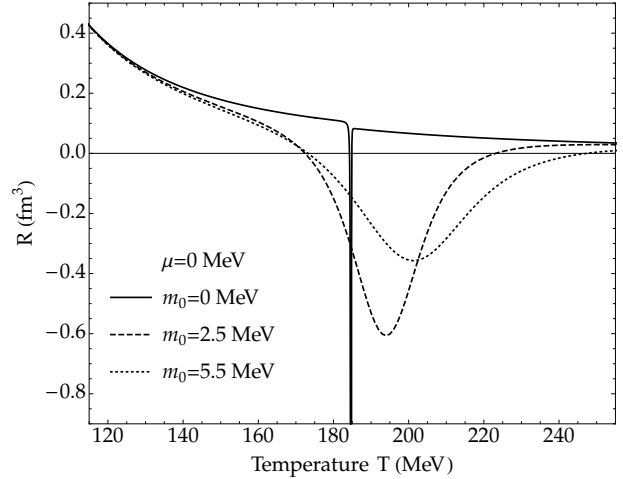}
	\caption{$R$ from $\mu=0$~MeV and different values of the bare mass $m_0$: continuous line is from $m_0=0$~MeV (the chiral limit) and $R$ shows a negative divergence. Dashed line is from $m_0=2.5$~MeV and the dotted from $m_0=5.5$~MeV; both show a finite region with negative $R$ around the transition temperature, which corresponds to the local maximum of $|R|$. }
	\label{fig:R2f}
\end{figure}

To clarify the effect of the  chiral mass in the calculation of the scalar curvature,  Fig.~\ref{fig:R2f} shows that $R$ diverges in the chiral limit but for  $m_0 \neq 0$, near the transition temperature,
it has a minimum, corresponding to a maximum of $|R|$, i.e. to a finite correlation length.
Therefore, $m_0 \neq 0$ changes the behavior of $R$ near the critical temperature: the divergence of the II order phase transition turns into  a minimum in the negative $R$ region 
and the transition temperature evaluated by the maximum of $|R|$ is completely in agreement with that one obtained by chiral susceptibility (see eq.~\eqref{eq:chi} in appendix~\ref{app:NJL2f}).

For low temperature and large chemical potential, the scalar curvature  $R$ has the same behavior previously discussed in the chiral limit, i.e. a first order phase transition. 

The critical point, $\left(T^\star, \mu^\star\right)$ between the crossover and the first order phase transition  depends on $m_0$ and for (the generally accepted value) $m_0=5.5$~MeV one has $\mu^\star\simeq 329$~MeV and $T^\star\sim 32$~MeV.

\begin{figure}
	\centering 
	\includegraphics[width=\columnwidth]{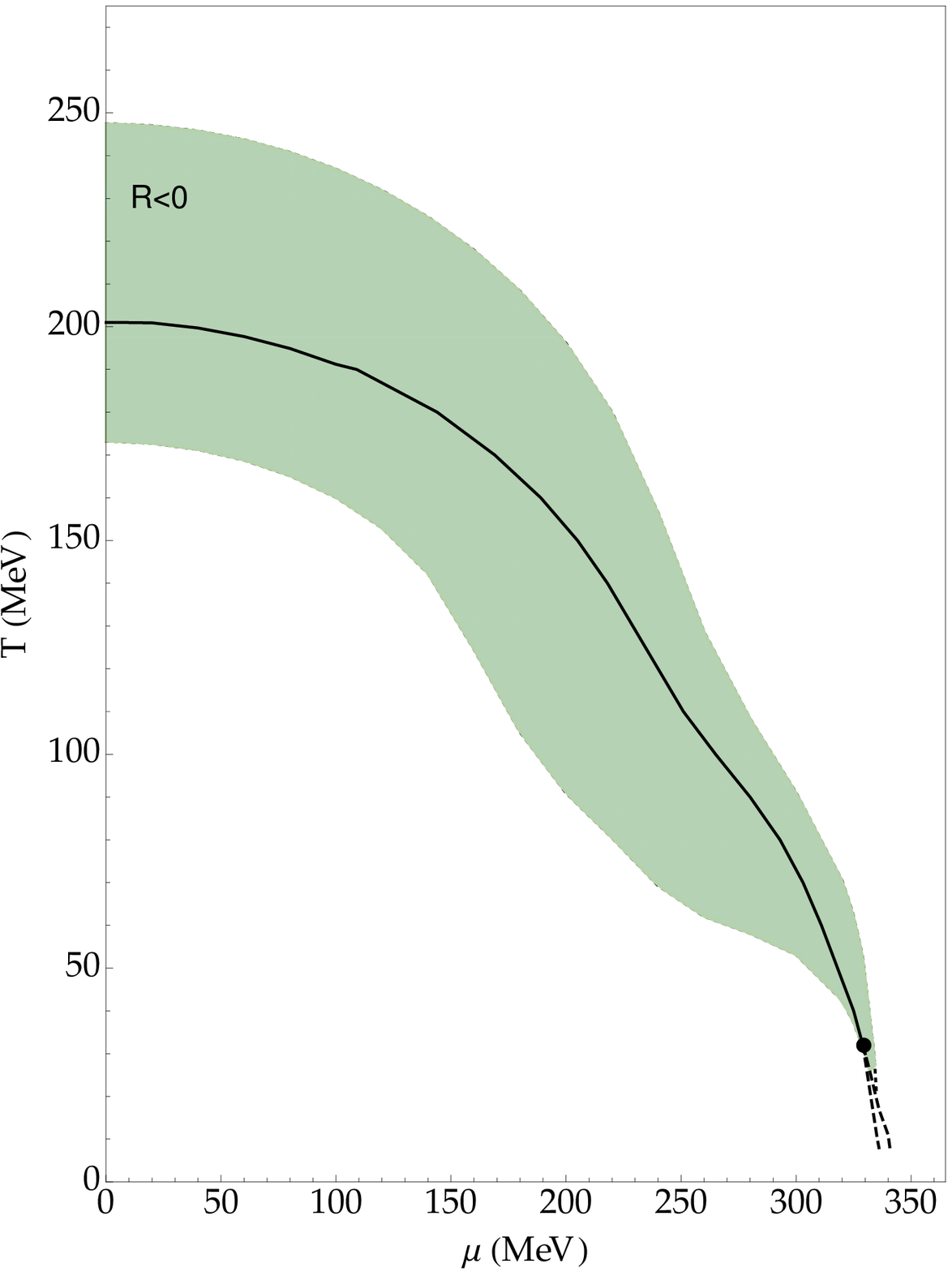}
	\caption{The transition temperature by the $R$ conditions and from $m_0=5.5$~MeV: continuous line is obtained by the local maximum of $|R|$, the dashed ones indicate the spinodal lines. The circle is at $\mu^\star=  329$~MeV and $T^\star=32$~MeV. 	The green band is the region of $R<0$.   }
	\label{fig:T2f}
\end{figure}

Figure~\ref{fig:T2f} shows the critical line for $m_0=5.5$~MeV: the continuous line is obtained by the maximum of $|R|$ and the dashed ones are the spinodal curves. The black circle is at $\mu^\star=  329$~MeV and $T^\star=32$~MeV. The green band is the region of $R<0$.

\subsubsection{Three flavors}
Three flavor NJL model is studied with the parameter values~\cite{Casalbuoni2005}
\begin{equation}
\begin{split}
\Lambda = 631.4\;\text{MeV}
&\quad\quad 
G\,\Lambda^2=1.835
\\
&K\,\Lambda^5 = 9.29
\\
m=5.5\;\text{MeV}
&\quad\quad 
m_s=135.7\;\text{MeV}
\end{split}
\end{equation}
and only one chemical potential ($\mu=\mu_d=\mu_u$, $\mu_s=0$). The dynamically generated masses $M_u=M_d$ and $M_s$ are now solutions of the system of eq.~\eqref{eq:GAP3q} and eq.~\eqref{eq:c}. Their behavior is similar to that one depicted in fig.~\ref{fig:M2f}, but with different values for light and strange quarks. Also in this case there is a crossover at low chemical potential and large $T$ and a first order phase transition at low temperature and large $\mu$.
The behavior of the scalar curvature is essentially the same of the previous case with two flavors and  physical masses. 

\begin{figure}
	\centering 
	\includegraphics[width=\columnwidth]{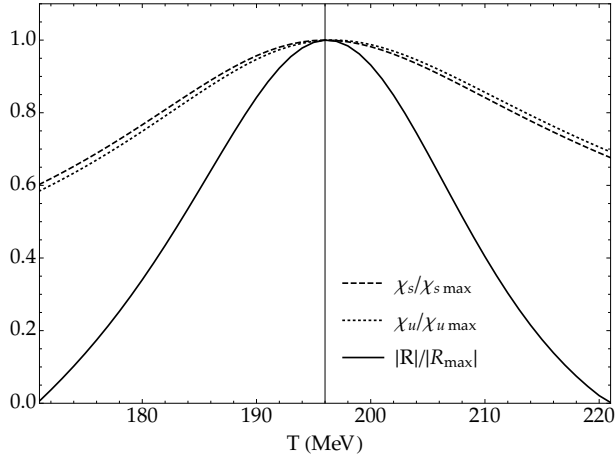}
	\caption{The ratio $\chi_s/\chi_{s max}$ (dashed line),  $\chi_u/\chi_{u max}$ (dotted line) and  $|R|/|R|_{max}$ (continuous line) at $\mu=0$~MeV.}
	\label{fig:NJLRchi}
\end{figure}

In figure~\ref{fig:NJLRchi}  the ratios $\chi_s/\chi_{s max}$ (dashed line), $\chi_u/\chi_{u max}$ (dotted line) and  $|R|/|R|_{max}$ (continuous line) are depicted to visualize that the maximum in $|R|$  corresponds to the peak of chiral susceptibilities. 

\begin{figure}
	\centering 
	\includegraphics[width=\columnwidth]{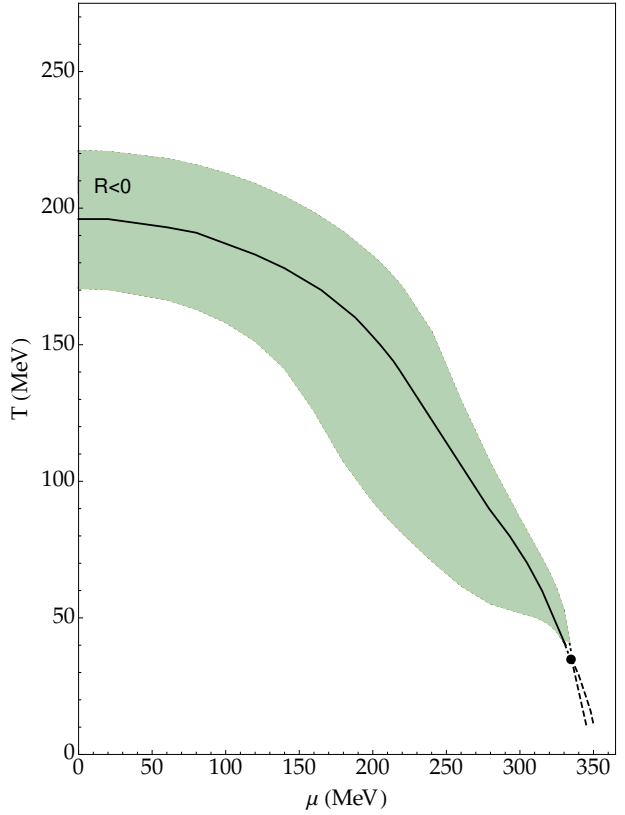}
	\caption{The transition temperature by the $R$ conditions: continuous line is obtained by the local maximum of $|R|$, the dashed ones indicate the spinodal lines. The circle is at $\mu^\star\sim  335$~MeV and $T^\star \sim 35$~MeV.The green band is the region of $R<0$ }
	\label{fig:T3f}
\end{figure}

Figure~\ref{fig:T3f} shows the transition temperature by the evaluation of $R$: the continuous line is again obtained by the maximum of $|R|$ and the dashed ones are the spinodal curves. The black circle is at $\mu^\star\sim335$~MeV and $T^\star\sim35$~MeV. The green band is the region of negative $R$.

\subsection{ Thermal geometric definition of the phase transitions in NJL model: summary}

It is useful to conclude this section by summarizing the geometrical definition of the phase transitions:
 
\begin{itemize}
	\item a II order phase transition occurs for two flavors  in the chiral limit ($m=0$) at low chemical potential. This transition is characterized by a divergent scalar curvature;

	\item for chiral masses, there is a crossover, both for two and three flavors, at low chemical potential and large $T$. 
              The transition temperature is defined as the maximum of $|R|$ in the negative-$R$ region and it is in agreement with the chiral susceptibility analysis $\chi$~\cite{Zhao:2008} (eqs.~\eqref{eq:chi}, \eqref{eq:chiu} and \eqref{eq:chis} in appendix);
	
	\item  there exists a I order phase transition at low temperature and large $\mu$, both with two and three flavors and both in the chiral limit or with chiral	 masses. This transition is related with  a discontinuity in $R$.

\end{itemize}

\section{NJL model and QCD crossover}\label{sec:QCD}

As seen in Sec.~\ref{sec:NJLG}, the NJL crossover  can be identified by a local maximum of $|R|$. 
However, NJL model misses color confinement and therefore there is no a priori reason to apply the same geometric criterium for non perturbative QCD dynamics. 

\begin{figure}
	\centering 
	\includegraphics[width=\columnwidth]{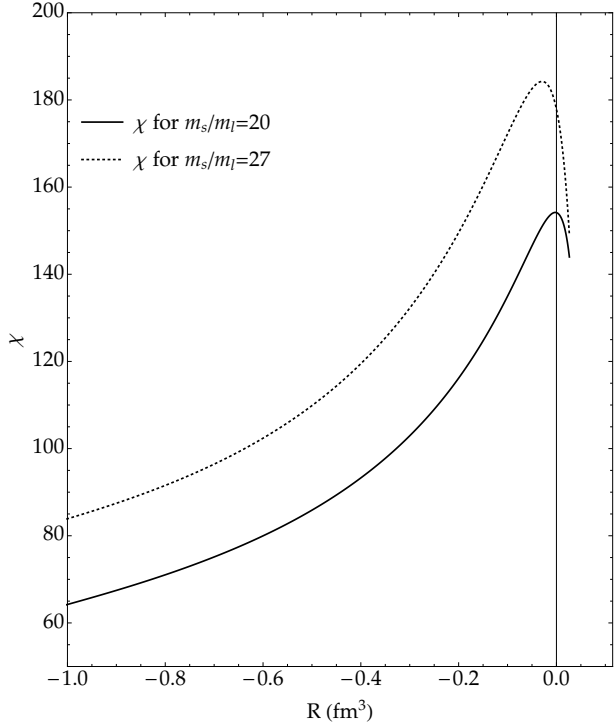}
	\caption{The chiral susceptibility $\chi$ at $\mu=0$~MeV and as a function of the scalar curvature $R$ for physical value of the strange quark mass, $m_s$, and $m_s/m_\ell = 20$ (dotted line) or $m_s/m_\ell = 27$ (continuous line).}
	\label{fig:ChiVsR}
\end{figure}

Indeed, in the thermodynamic geometry description of QCD deconfinement transition  in  ref.~\cite{Castorina:2018ayy} the criterium $R=0$ has been used. It indicates the transition from a mostly fermionic system (as the quark-gluon plasma) to an essentially bosonic one (as the hadron resonance gas) but, as shown in fig.~\ref{fig:ChiVsR}, it exactly corresponds to the maximum of chiral susceptibility, 
confirming the well known \cite{Casher:1979vw,Banks:1979yr,Castorina:1981iy,Digal:2000ar}
interplay between confinement and chiral symmetry breaking.

The transition temperature evaluated by \mbox{$R=0$} is  in agreement with the freeze-out hadronization curve and with the pseudo-critical temperature by lattice data within $10\%$ \cite{Castorina:2018ayy}.

\section{Comments and Conclusions}\label{sec:CC}

Thermodynamic geometry reliably describes the phase diagram of NJL model, both in the chiral limit and for finite mass, and indicates a geometrical interplay between chiral symmetry restoration/breaking and deconfinement/confinement regimes.   

Moreover in a very recent paper \cite{Ding:2018auz}
the chiral phase transition temperature $T^0_c$, corresponding to a ``true'' chiral transition in the limit \mbox{$m_s/m_l >> 1$}, 
turns out to be about $25$ MeV less than the pseudo-critical temperature.

Fig~\ref{fig:ChiVsR} suggests that a small variation from $m_s/m_l = 20 $  to $m_s/m_l = 27 $ changes the maximum of chiral susceptibility  from  $R=0$ to a finite value of $|R|$, as in NJL model.
It could be possible that considering the effective chiral limit, i.e. $m_s/m_l >> 1$ one recovers by thermodynamic geometry a ``true'' chiral phase transition at lower temperature, with typical scaling laws.
The role of color confinement in QCD in terms of thermodynamic geometry will be discussed in different models in  a forthcoming paper.
\vskip 10pt
{\bf Acknowledgements}  
The authors thank H.Satz for useful comments.

\appendix

\section{NJL model with two quarks\label{app:NJL2f}}
To evaluate the scalar curvature $R$ one needs the derivatives of the potential $\phi$, up to third order, which can be written in terms of the dynamical generated mass $M$. Therefore, the solution of the GAP equation uniquely determines all those functions. Indeed, after a straightforward calculation, one gets (a comma indicates partial derivative)
\begin{equation}\label{eq:Mb}
M_{,\beta} =\frac{b_1\;M}{1-f_1-f_2\;M^2} \;,
\end{equation}
\begin{equation}
M_{,\gamma} =\frac{g_1\;M}{1-f_1-f_2\;M^2}\;,
\end{equation}
\begin{equation}\label{eq:Mbb}
\begin{split}
M_{,\beta\beta} =&
d\;\Bigg[b_3\;M+\left(b_2+f_{1,\beta}\right)M_{,\beta} +\\
&+ \left(b_4+2\,T\,f_2\right)M^2\,M_{,\beta}+f_3\;M\,M_{,\beta}^2\Bigg] \;,
\end{split}
\end{equation}
\begin{equation}
\begin{split}
M_{,\gamma\gamma} =&
d\;\Bigg[g_3\;M+\left(g_2+f_{1,\gamma}\right)M_{,\gamma} + g_4\,M^2\,M_{,\gamma}+\\
&+f_3\;M\,M_{,\gamma}^2\Bigg] \;,
\end{split}
\end{equation}
\begin{equation}
\begin{split}
M_{,\beta\gamma} 
=&
d\;\Bigg[g_5\;M+b_2\,M_{,\gamma} + f_{1,\gamma} M_{,\beta} + g_4\,M^2\,M_{,\beta}+\\
&+f_2\;T\,M^2\,M_{,\gamma}+f_3\;M\,M_{,\beta} M_{,\gamma}\Bigg] \;,
\end{split}
\end{equation}
with
\begin{equation}
d = \left(1-f_1-f_2\;M^2\right)^{-1}
\;,
\end{equation}
\begin{equation}
f_{1} =
\kappa_M\;
\int_0^\Lambda dp\;\frac{p^4\;\Psi}{E^3}
\;,
\end{equation}
\begin{equation}
f_2=
\kappa_M\,\int_0^\Lambda dp\;p^2\;\frac{n_-\left(1-n_-\right)+n_+ \left(1-n_+\right)}{T\;E^2}\;,
\end{equation} 
\begin{equation}
f_3=
\kappa_M\,\int_0^\Lambda dp\;p^4\;\frac{n_-\left(1-n_-\right)+n_+ \left(1-n_+\right)}{T\;E^4}\;,
\end{equation} 
\begin{equation}
b_1 =
\kappa_M\,\int_0^\Lambda dp\;p^2\;\left[n_-\left(1-n_-\right)+n_+ \left(1-n_+\right)\right]\;,
\end{equation}
\begin{equation}
b_2 =
\kappa_M\,\int_0^\Lambda dp\;p^4\;\frac{\Psi_{\beta}}{E^3} \;,
\end{equation}
\begin{equation}
b_3\!=\!
\kappa_M\!\!\int_0^\Lambda\!\!\! dp p^2\;\left[n_{-,\beta}\left(1-2n_-\right)+n_{+,\beta} \left(1-2n_+\right)\right]\;,
\end{equation}
\begin{equation}
b_4\!=\!
\kappa_M\!\!\int_0^\Lambda\!\!\! dp p^2\;\frac{n_{-,\beta}\left(1-2n_-\right)+n_{+,\beta} \left(1-2n_+\right)}{T\;E^2}\;,
\end{equation} 
\begin{equation}
g_1 =
\kappa_M\,\int_0^\Lambda dp\;p^2\;\frac{n_- \left(1-n_-\right) - n_+\left(1-n_+\right)}{E}\;,
\end{equation}
\begin{equation}
g_2 =
\kappa_M\,\int_0^\Lambda dp\;p^4\;\frac{\Psi_{\gamma}}{E^3} \;,
\end{equation}
\begin{equation}
g_3\!=\!
\kappa_M\!\!\int_0^\Lambda\!\!\! dp p^2\;\frac{n_{-,\gamma}\left(1-2n_-\right)-n_{+,\gamma} \left(1-2n_+\right)}{E}\;,
\end{equation}
\begin{equation}
g_4\!=\!
\kappa_M\!\!\int_0^\Lambda\!\!\! dp p^2\;\frac{n_{-,\gamma}\left(1-2n_-\right)+n_{+,\gamma} \left(1-2n_+\right)}{T\;E^2}\;,
\end{equation} 
\begin{equation}
g_5\!=\!
\kappa_M\!\!\int_0^\Lambda\!\!\! dp p^2\;\left[n_{-,\gamma}\left(1-2n_-\right)+n_{+,\gamma} \left(1-2n_+\right)\right]\;,
\end{equation}
\begin{equation}
\kappa_M = 2\;G\;\frac{N_c\,N_f}{\pi^2}
\end{equation}
and $n{\pm}$ in Eq.~\eqref{eq:n}.

By deriving Eq.~\eqref{eq:phi} and Eq.~\eqref{eq:Omega} and defining 
\begin{equation}
\kappa_\Omega = \frac{\kappa_M}{2\,G} \;,
\end{equation}
one gets
\begin{equation}
\phi_{,\beta} =
\kappa_\Omega \int_0^\Lambda dp\,p^2\,E\,\Psi  -\frac{\left(M-m\right)^2}{4\,G}\;,
\end{equation}
\begin{equation}
\phi_{,\gamma} =
\kappa_\Omega \int_0^\Lambda dp\,p^2\,\left(n_+ - n_-\right)  \;.
\end{equation}
The calculation of second and third order derivatives is straightforward. 

Finally, the two flavors chiral susceptibility, $\chi$, is defined as~\cite{Zhao:2008}
\begin{equation}\label{eq:chi}
\begin{split}
\chi^{2f} =
\frac{\partial M}{\partial m} 
=& \frac{1}{1 - f_1
	-	f_2\;M^2} 	=\\
=&
\frac{M_{,\beta}}{b_1\;M} 
=
\frac{M_{,\gamma}}{g_1\;M} 
\;.
\end{split}
\end{equation}

\section{Three flavors\label{app:NJL3f}}
In a three flavors systems the derivatives of the dynamically generated mass $M_u=M_d$ and $M_s$ are
\begin{equation}
\begin{cases}
\displaystyle
M_{u,\beta}\left(\delta - b_u\,M_u\,\epsilon\right) = 
a_u\,\epsilon - M_{s,\beta}\,\zeta
\\\\
\displaystyle
M_{s,\beta}
=
\frac{\left(a_s\,\theta - a_u\,\lambda\right)\left(\delta - b_u\,M_u\,\epsilon\right) 
	-
	a_u\,\epsilon\,b_u\,\lambda \,M_u}
{\left(\
	\eta -b_s\,M_s\,\theta \right) 
	\left(\delta - b_u\,M_u\,\epsilon\right)-
	b_u\,\lambda \,M_u\,\zeta}
\end{cases}
,
\end{equation}
\begin{equation}
\begin{cases}
\displaystyle
M_{u,\gamma}\left(\delta - b_u\,M_u\,\epsilon\right) = 
c_u\,\epsilon - M_{s,\gamma}\,\zeta
\\\\
\displaystyle
M_{s,\gamma}
=
\frac{\left(c_s\,\theta - c_u\,\lambda\right)\left(\delta - b_u\,M_u\,\epsilon\right) 
	-
	c_u\,\epsilon\,b_u\,\lambda \,M_u}
{\left(\
	\eta -b_s\,M_s\,\theta \right) 
	\left(\delta - b_u\,M_u\,\epsilon\right)-
	b_u\,\lambda \,M_u\,\zeta}
\end{cases}
,
\end{equation}
\begin{widetext}
\begin{equation}
\begin{cases}
\displaystyle
M_{u,\beta\beta}\left(\delta - b_u\,M_u\,\epsilon\right) = 
d_u\,\epsilon + A_{u,\beta}\,\epsilon_{,\beta}-
\left(M_{u,\beta}\delta_{,\beta}+M_{s,\beta}\zeta_{,\beta}+\epsilon\,D_{u,\beta}\,M_u\,M_{u,\beta}\right)
-
M_{s,\beta\beta}\,\zeta
\\\\
\displaystyle
\begin{split}
M_{s,\beta\beta}
=
&
\frac{\left(d_s\,\theta-d_u\,\lambda
	+
	A_{s,\beta}\,\theta_{,\beta}-A_{u,\beta}\,\lambda_{,\beta}
	+
	\lambda\,D_{u,\beta}\,M_u\,M_{u,\beta}
	-
	\theta\,D_{s,\beta}\,M_s\,M_{s,\beta}\right)\left(\delta - b_u\,M_u\,\epsilon\right)}{
	\left(\eta - b_s\,M_s\,\theta\right)\left(\delta - b_u\,M_u\,\epsilon\right)-\zeta\,\lambda\,b_u\,M_u}
-\\
&-
\lambda\,b_u\,M_u\,\frac{\left[d_u\,\epsilon + A_{u,\beta}\,\epsilon_{,\beta}-
	\left(M_{u,\beta}\delta_{,\beta}+M_{s,\beta}\zeta_{,\beta}+\epsilon\,D_{u,\beta}\,M_u\,M_{u,\beta}\right)\right]}{
	\left(\eta - b_s\,M_s\,\theta\right)\left(\delta - b_u\,M_u\,\epsilon\right)-\zeta\,\lambda\,b_u\,M_u }
\end{split} 
\end{cases}
,
\end{equation}
\begin{equation}
\begin{cases}
\displaystyle
M_{u,\gamma\gamma}\left(\delta - b_u\,M_u\,\epsilon\right) = 
e_u\,\epsilon + A_{u,\gamma}\,\epsilon_{,\gamma}-
\left(M_{u,\gamma}\delta_{,\gamma}+M_{s,\gamma}\zeta_{,\gamma}+\epsilon\,D_{u,\gamma}\,M_u\,M_{u,\gamma}\right)
-
M_{s,\gamma\gamma}\,\zeta
\\\\
\displaystyle
\begin{split}
M_{s,\gamma\gamma}
=
&
\frac{\left(e_s\,\theta-e_u\,\lambda
	+
	A_{s,\gamma}\,\theta_{,\gamma}-A_{u,\gamma}\,\lambda_{,\gamma}
	+
	\lambda\,D_{u,\gamma}\,M_u\,M_{u,\gamma}
	-
	\theta\,D_{s,\gamma}\,M_s\,M_{s,\gamma}\right)\left(\delta - b_u\,M_u\,\epsilon\right)}{
	\left(\eta - b_s\,M_s\,\theta\right)\left(\delta - b_u\,M_u\,\epsilon\right)-\zeta\,\lambda\,b_u\,M_u}
-\\
&-
\lambda\,b_u\,M_u\,\frac{\left[e_u\,\epsilon + A_{u,\gamma}\,\epsilon_{,\gamma}-
	\left(M_{u,\gamma}\delta_{,\gamma}+M_{s,\gamma}\zeta_{,\gamma}+\epsilon\,D_{u,\gamma}\,M_u\,M_{u,\gamma}\right)\right]}{
	\left(\eta - b_s\,M_s\,\theta\right)\left(\delta - b_u\,M_u\,\epsilon\right)-\zeta\,\lambda\,b_u\,M_u }
\end{split} 
\end{cases}
\end{equation}
and
\begin{equation}
\begin{cases}
\displaystyle
M_{u,\beta\gamma}\left(\delta - b_u\,M_u\,\epsilon\right) = 
f_u\,\epsilon + A_{u,\beta}\,\epsilon_{,\gamma}-
\left(M_{u,\beta}\delta_{,\gamma}+M_{s,\beta}\zeta_{,\gamma}+\epsilon\,D_{u,\beta}\,M_u\,M_{u,\gamma}\right)
-
M_{s,\beta\gamma}\,\zeta
\\\\
\displaystyle
\begin{split}
M_{s,\beta\gamma}
=
&
\frac{\left(f_s\,\theta-f_u\,\lambda
	+
	A_{s,\beta}\,\theta_{,\gamma}-A_{u,\beta}\,\lambda_{,\gamma}
	+
	\lambda\,D_{u,\beta}\,M_u\,M_{u,\gamma}
	-
	\theta\,D_{s,\beta}\,M_s\,M_{s,\gamma}\right)\left(\delta - b_u\,M_u\,\epsilon\right)}{
	\left(\eta - b_s\,M_s\,\theta\right)\left(\delta - b_u\,M_u\,\epsilon\right)-\zeta\,\lambda\,b_u\,M_u}
-\\
&-
\lambda\,b_u\,M_u\,\frac{\left[f_u\,\epsilon + A_{u,\beta}\,\epsilon_{,\gamma}-
	\left(M_{u,\beta}\delta_{,\gamma}+M_{s,\beta}\zeta_{,\gamma}+\epsilon\,D_{u,\beta}\,M_u\,M_{u,\gamma}\right)\right]}{
	\left(\eta - b_s\,M_s\,\theta\right)\left(\delta - b_u\,M_u\,\epsilon\right)-\zeta\,\lambda\,b_u\,M_u }
\end{split} 
\end{cases}
,
\end{equation}
\end{widetext} 
where 
\begin{equation}
a_f
= \frac{N_c}{\pi^2} 
\int_0^\Lambda dp\, p^2
\left[
n_{-f}\left(1-n_{-f} \right)
+
n_{+f}\left(1-n_{+f} \right) 
\right]\;,
\end{equation}
\begin{equation}
b_f
= \frac{N_c}{\pi^2} 
\int_0^\Lambda dp\, p^2
\frac{ 
	n_{-f}\left(1-n_{-f} \right)
	+
	n_{+f}\left(1-n_{+f} \right) 
}{T\;E^2_f}\;,
\end{equation}
\begin{equation}
c_f
= \frac{N_c}{\pi^2} 
\int_0^\Lambda dp\, p^2
\frac{ 
	n_{-f}\left(1-n_{-f} \right)
	-
	n_{+f}\left(1-n_{+f} \right) 
}{E_f}\;,
\end{equation}
\begin{widetext}
\begin{equation}
\begin{split}
d_f
= \frac{N_c}{\pi^2} 
\int_0^\Lambda dp\; p^2\Bigg\{&
\left(\frac{2\,M\,M_{,\beta}}{E^2} +p^2\;\frac{M^2_{,\beta}}{T\,E^4}\right)\left[ 
n_{-f}\left(1-n_{-f} \right)
+
n_{+f}\left(1-n_{+f} \right) \right] 
+\\
&+
\left(1 +\frac{M\,M_{,\beta}}{T\,E^2}\right)\left[ 
\left(1-2\,n_{-f} \right)n_{-f,\beta}
+
\left(1-2\,n_{+f} \right)n_{+f,\beta} 
\right] \Bigg\}\;,
\end{split}
\end{equation}
\begin{equation}
\begin{split}
e_f
= \frac{N_c}{\pi^2} 
\int_0^\Lambda dp\; p^2\Bigg\{&
p^2\;\frac{M^2_{,\gamma}}{T\,E^4}\left[ 
n_{-f}\left(1-n_{-f} \right)
+
n_{+f}\left(1-n_{+f} \right) \right] 
+\\
&+
\frac{M\,M_{,\gamma}}{T\,E^2}\left[ 
\left(1-2\,n_{-f} \right)n_{-f,\gamma}
+
\left(1-2\,n_{+f} \right)n_{+f,\gamma} 
\right] +\\
&+
\frac{\left(1-2\,n_{-f} \right)n_{-f,\gamma}
	-
	\left(1-2\,n_{+f} \right)n_{+f,\gamma} }{E} 
\Bigg\}\;,
\end{split}
\end{equation}
\begin{equation}
\begin{split}
f_f
= \frac{N_c}{\pi^2} 
\int_0^\Lambda dp\; p^2\Bigg\{&
\left(\frac{M\,M_{,\gamma}}{E^2} +p^2\;\frac{M_{,\beta}M_{,\gamma}}{T\,E^4}\right)\left[
n_{-f}\left(1-n_{-f} \right)
+
n_{+f}\left(1-n_{+f} \right) \right] 
+\\
&+
\left(1 +\frac{M\,M_{,\beta}}{T\,E^2}\right)\left[ 
\left(1-2\,n_{-f} \right)n_{-f,\gamma}
+
\left(1-2\,n_{+f} \right)n_{+f,\gamma} 
\right] \Bigg\}\;,
\end{split}
\end{equation}
\end{widetext}
\begin{equation}
A_{f,\beta}=a_f + b_f\;M_f\,M_{f,\beta}\;,
\end{equation}
\begin{equation}
A_{f,\gamma}=c_f + b_f\;M_f\,M_{f,\gamma}\;,
\end{equation}
\begin{equation}
C_{f,\beta\beta}=d_f + b_f\;M_f\,M_{f,\beta\beta}\;,
\end{equation}
\begin{equation}
C_{f,\gamma\gamma}=e_f + b_f\;M_f\,M_{f,\gamma\gamma}\;,
\end{equation}
\begin{equation}
C_{f,\beta\gamma}=f_f + b_f\;M_f\,M_{f,\beta\gamma}\;,
\end{equation}
\begin{equation}
B(M_u,M_s) = 4\,G-2\,\frac{K^2}{G}\;u^2-2\,K\,s \;,
\end{equation}
\begin{equation}
\delta(M_u,M_s)
=
\left(1 - F_{1u}\,B\right)\;,
\end{equation}
\begin{equation}
\zeta(M_u,M_s) = 2\,K\,u\;,
\end{equation}
\begin{equation}
\epsilon(M_u,M_s)=
B\,M_u\;,
\end{equation}
\begin{equation}
\begin{split}
\eta(M_u,M_s)=&\left(1-4GF_{1s}\right)\left(1-F_{1u}B\right)-\\
& 
-8K^2u^2F_{1u}\;,
\end{split}
\end{equation}
\begin{equation}
\theta(M_u,M_s)
=4\,G\,\left(1-F_{1u}B\right)M_s\,,
\end{equation}
\begin{equation}
\lambda(M_u,M_s)
=
4\,K\,u\,M_u\;,
\end{equation}
\begin{equation}
F_{1f}
=
\frac{N_c}{\pi^2}\int_0^\Lambda dp\,p^4\,\frac{ \Psi_{f}}{E^3_f}
\;,
\end{equation}
\begin{equation}
n_{f\pm}
=\frac{1}{1+\exp\left\{ \frac{\sqrt{p^2 + M^2_f} \pm \mu_f}{T} \right\}} 
\end{equation}
and $u\equiv \left<\overline u u\right>$, $s\equiv \left<\overline s s\right>$.

About the thermodynamic potential $\phi=-\Omega\,\beta$ one has
\begin{equation}
\begin{split}
\phi_{,\beta} =
\sum_{f=u,d,s}&
\frac{N_c}{\pi^2}\int^\Lambda_0 dp\,p^2\,E_f\,\Psi_f
+2\,G\,s^2+\\
& + u\,\left(M_u-m_u\right)+s\,\left(M_s-m_s\right)=\\
\sum_{f=u,d,s}&
\frac{N_c}{\pi^2}\int^\Lambda_0 dp\,p^2\,E_f\,\Psi_f
+K\,u^2\,s+\\
& + u\,\left(M_u-m_u\right)+\frac{s\,\left(M_s-m_s\right)}{2} \;,
\end{split}
\end{equation}
\begin{equation}
\phi_{,\gamma} =
\sum_{f=u,d,s}
\frac{N_c}{\pi^2}\int^\Lambda_0 dp\,p^2\,\left(n_{+f}-n_{-f}\right)
\end{equation}
\begin{equation}
\phi_{,\beta\beta} \!\!=\!\!
-\!\!\!\!\sum_{f=u,d,s}\!\!\!\!
\frac{N_c}{\pi^2}\int^\Lambda_0\!\!\! dp\,p^2E_f\left(n_{+f,\beta}+n_{-f,\beta}\right)
\end{equation}
\begin{equation}
\phi_{,\beta\gamma} =
\sum_{f=u,d,s}
\frac{N_c}{\pi^2}\int^\Lambda_0 dp\,p^2\,\left(n_{+f,\beta}-n_{-f,\beta}\right)
\end{equation}
\begin{equation}
\phi_{,\gamma\gamma} =
\sum_{f=u,d,s}
\frac{N_c}{\pi^2}\int^\Lambda_0 dp\,p^2\,\left(n_{+f,\gamma}-n_{-f,\gamma}\right)
\end{equation}
\begin{equation}
\phi_{,\beta\beta\gamma} =
\sum_{f=u,d,s}
\frac{N_c}{\pi^2}\int^\Lambda_0 dp\,p^2\,\left(n_{+f,\beta\beta}-n_{-f,\beta\beta}\right)
\end{equation}
\begin{equation}
\phi_{,\beta\gamma\gamma} =
\sum_{f=u,d,s}
\frac{N_c}{\pi^2}\int^\Lambda_0 dp\,p^2\,\left(n_{+f,\beta\gamma}-n_{-f,\beta\gamma}\right)
\end{equation}
\begin{equation}
\phi_{,\gamma\gamma\gamma} =
\sum_{f=u,d,s}
\frac{N_c}{\pi^2}\int^\Lambda_0 dp\,p^2\,\left(n_{+f,\gamma\gamma}-n_{-f,\gamma\gamma}\right)
\end{equation}
\begin{equation}
\begin{split}
\phi_{,\beta\beta\beta} \!\!=\!\!&
-\!\!\!\!\sum_{f=u,d,s}\!\!\!\!
\frac{N_c}{\pi^2}\int^\Lambda_0\!\!\! dp\,p^2E_f\left(n_{+f,\beta\beta}+n_{-f,\beta\beta}\right)
+\\
&+\!\!\!\!
\sum_{f=u,d,s}\!\!\!\!\left( a_f + b_f\,M_f\,M_{f,\beta}\right) M_f\,M_{f,\beta}
\end{split}
\end{equation}
Finally, by defining
\begin{equation}
H_f = F_{1f} + b_f\;M^2_f
\;,
\end{equation}
the chiral susceptibilities are
\begin{widetext}
\begin{equation}\label{eq:chiu}
\chi_u=\chi_d = 
\frac{\partial M_u}{\partial m_u} =
\frac{1 - 4\,G\,H_s}{1 - 4\,G(H_u+H_s)+4\,H_u\,H_s(4\,G^2-2\,K\,G\,s-K^2\,u^2)+2\,K\,s\,H_u}
\end{equation}
	and
	\begin{equation}\label{eq:chis}
	\chi_s = 
	\frac{\partial M_s}{\partial m_s} =
		=
	\frac{1 - (4\,G-2\,K\,s)\,H_u}{1 - 4\,G(H_u+H_s)+4\,H_u\,H_s(4\,G^2-2\,K\,G\,s-2\,K^2\,u^2)+2\,K\,s\,H_u} 
	\end{equation}

\end{widetext}

\end{document}